\documentclass[12pt,preprint]{emulateapj}
\usepackage{color}
\usepackage{url}

\begin{document}

\title{An ALMA Disk Mass for the Candidate Protoplanetary Companion to FW Tau}

\author{
Adam L. Kraus\altaffilmark{1},
Sean M. Andrews\altaffilmark{2},
Brendan P. Bowler\altaffilmark{3,4},
Gregory Herczeg\altaffilmark{5},
Michael J. Ireland\altaffilmark{6}, \\
Michael C. Liu\altaffilmark{7},
Stanimir Metchev\altaffilmark{8},
Kelle L. Cruz\altaffilmark{9}
}

\altaffiltext{1}{Department of Astronomy, The University of Texas at Austin, Austin, TX 78712, USA}
\altaffiltext{2}{Harvard-Smithsonian Center for Astrophysics, 60 Garden Street, Cambridge, MA 02138, USA}
\altaffiltext{3}{California Institute of Technology, Division of Geological and Planetary Sciences, 1200 East California Boulevard, Pasadena, CA 91101, USA}
\altaffiltext{4}{Caltech Joint Center for Planetary Astronomy Fellow}
\altaffiltext{5}{Kavli Institute for Astronomy and Astrophysics, Peking University, Yi He Yuan Lu 5, Haidian Qu, Beijing 100871, China}
\altaffiltext{6}{Research School of Astronomy \& Astrophysics, Australian National University, Canberra ACT 2611, Australia}
\altaffiltext{7}{Institute for Astronomy, University of Hawai'i, 2680 Woodlawn Drive, Honolulu, HI 96822, USA}
\altaffiltext{8}{Department of Physics and Astronomy, The University of Western Ontario, London, ON N6A 3K7, Canada}
\altaffiltext{9}{Department of Physics and Astronomy, Hunter College, City University of New York, New York, NY 10065, USA}

\begin{abstract}

We present ALMA observations of the FW Tau system, a close binary pair of M5 stars with a wide-orbit (300\,AU projected separation) substellar companion. The companion is extremely faint and red in the optical and near-infrared, but boasts a weak far-infrared excess and optical/near-infrared emission lines indicative of a primordial accretion disk of gas and dust. The component-resolved 1.3\,mm continuum emission is found to be associated only with the companion, with a flux ($1.78\pm0.03$\,mJy) that indicates a dust mass of 1--2\,$M_{\oplus}$. While this mass reservoir is insufficient to form a giant planet, it is more than sufficient to produce an analog of the Kepler-42 exoplanetary system or the Galilean satellites. The mass and geometry of the disk-bearing FW Tau companion remains unclear. Near-infrared spectroscopy shows deep water bands that indicate a spectral type later than M5, but substantial veiling prevents a more accurate determination of the effective temperature (and hence mass).  Both a disk-bearing ``planetary-mass'' companion seen in direct light or a brown dwarf tertiary viewed in light scattered by an edge-on disk or envelope remain possibilities. 

\end{abstract}


\section{Introduction}

Over the past decade, direct imaging surveys have discovered a small but significant number of faint, apparently planetary-mass ($\la$20 $M_{Jup}$) companions that orbit their primary star hosts ($M \sim$0.2--1.5\,$M_{\odot}$) at ultrawide separations ($\ga$100\,AU, extending to thousands of AU) \citep{Neuhauser:2005ea,Lafreniere:2008oy,Schmidt:2008pd,Ireland:2011fj,Bailey:2014fk,Kraus:2014lr}. These planetary mass companions (PMCs) represent intriguing analogs to the recent discoveries of smaller separation planets, like HR 8799 bcde \citep{Marois:2008zt}, Beta Pic b \citep{Lagrange:2009fc}, HD 95086 b \citep{Rameau:2013gf}, GJ 504 b \citep{Kuzuhara:2013ly}, and LkCa15 b \citep{Kraus:2012nx}. The large orbital separations of the PMCs are markedly different from the planets in our own solar system and the vast population of exoplanets detected with the radial velocity and transit methods, so it is not clear whether PMCs formed via similar processes. Planets at orbital radii of $\la$100 AU can feasibly be formed via traditional methods like core accretion \citep{Pollack:1996dk} and Class II disk instability \citep{Boss:2011fj}. At wider radii, the most plausible process is likely disk fragmentation at the Class 0/I stage\citep[e.g.,][]{Kratter:2010yr}. Nonetheless, PMCs are a potential boon for exoplanet studies; their large separations make them relatively easy to observe, so they could serve as the fully characterized templates against which the more difficult measurements of ``traditional" planets are compared.

In particular, PMCs offer an unique window into the process of giant planet assembly and the associated formation of moon systems. Most PMCs have been found in star-forming regions with ages $\la 10$\,Myr, where gas-rich protoplanetary disks are common. Several PMCs exhibit emission lines or possible mid-infrared excesses \citep{Seifahrt:2007ve,Schmidt:2008pd,Bowler:2011kx,Bailey:2013fj,Bowler:2014lr} that are commonly associated with disks and outflows. Recent {\it Hubble Space Telescope} ({\it HST}) data confirm that some PMCs have large optical/ultraviolet excess emission, indicative of shocks due to the accretion of disk material \citep[][]{Zhou:2014fk}. Such observations demonstrate that PMCs can host their own circum(sub)stellar disks, composed of material left over from their own formation or accreted from the disk/envelope of their (much more massive) host. 

Measurements of the masses, structures, and lifetimes of these disks provide constraints on the assembly timescale for PMCs and the duration of their satellite formation epoch which could be compared with measurements of the complementary free-floating substellar population; \citep[e.g.,][]{Liu:2003me,Scholz:2006lw,Andrews:2013uq}. Furthermore, hydrodynamic models of giant planet formation \citep[e.g.,][]{Ayliffe:2009aa} make predictions of the disk radii, scale heights, and mass distributions that could be tested with spatially resolved observations, and wide companions also provide context for observations of close-in planets with ring systems, such as the recently discovered companion to 1SWASP J140747.93-394542.6 \citep{Mamajek:2012pb,Kenworthy:2015lr}. However, these disks also pose challenges: if they are oriented at large viewing angles \citep[edge-on;][]{Scholz:2008fk,Luhman:2007kl,Looper:2010zr}, the disk material can obscure the PMC and complicate the determination of its mass.

\section{The FW Tau System}

FW Tau is a member of the Taurus-Auriga association, a nearby ($d \sim 140$ pc), young ($\tau \sim 2$ Myr) region of ongoing star formation. The well-studied primary consists of a close pair ($\sim$75\,mas, or 11\,AU, separation) of M5 stars (FW Tau AB) with a total mass of 0.2--0.3\,$M_{\odot}$ \citep{Baraffe:1998yo}. FW Tau AB exhibits no spectroscopic evidence of ongoing accretion \citep[][]{Bowler:2014lr}, nor any clear signature of an infrared excess shortward of 24\,$\mu$m \citep[][]{Luhman:2010cr}, indicating that it does not host a significant optically-thick disk at orbital radii $<$50\,AU. Intriguingly, \citet[][]{Andrews:2005qf} noted a 4$\sigma$ detection of the system at 850\,$\mu$m ($F_{\nu}=4.5 \pm 1.1$\,mJy), suggesting that a small amount of cool dust is present within a $\sim$15\arcsec diameter region around the system. Recent Herschel observations by \citet[][]{Howard:2013qy} also find a far-infrared excess within a similar (or larger) beam around FW Tau ($F_{\nu} = 30 \pm 4$, $33\pm4$, and $70\pm40$\,mJy at 70, 100, and 160\,$\mu$m, respectively).  

\citet[][]{White:2001jf} first reported an extremely faint candidate (tertiary) companion near FW Tau AB in {\it HST} optical images, with anomalously red $i'z'$ photometry and strong (narrowband) H$\alpha$ emission. \citet{Kraus:2014lr} subsequently confirmed that this companion (named either FW Tau C or FW Tau b, depending on its poorly-understood nature and whether it is best considered as a binary companion or planetary-type companion) and FW Tau AB were co-moving using ground-based adaptive optics imaging, and began a multi-pronged effort to better characterize its properties. If the companion's flux can be attributed to unobscured photospheric emission, the near-infrared flux suggests a companion mass as low as 10\,$M_{Jup}$, making it another example of a PMC. However, the true mass could be significantly higher or lower if there is an excess from accretion or disk emission (as for many disk-hosting stars and brown dwarfs; \citealt{Luhman:2010cr}) and/or obscuration from an envelope or edge-on disk (as for a number of other stellar or substellar companions; \citealt{Stapelfeldt:1998iv,Scholz:2008fk,Luhman:2007kl,Duchene:2010xt}). \citet{Bowler:2014lr} recently obtained a near-infrared spectrum of FW Tau C that confirms the previous hints of accretion signatures, but found that veiling obscures any photospheric features except broad water absorption bands, which are present for late-M or early-L dwarfs. The spectra therefore remain consistent with either a PMC or a brown dwarf obscured by an edge-on disk.

The tentative detection of submillimeter emission by \citet[][]{Andrews:2005qf} from the unresolved FW Tau system hinted that a disk could be present around at least one component, and the high H$\alpha$ line flux observed from the faint companion by \citet[][]{White:2001jf} strongly suggested that material was accreting onto the companion.  In this Letter, we present sensitive, component-resolved 1.3\,mm-wavelength observations from the Atacama Large Millimeter Array (ALMA) designed to confirm and localize the putative long-wavelength continuum emission (to FW Tau AB or to FW Tau C) and to more accurately measure the mass of the dust responsible for it.

\section{ALMA Observations and Data Reduction}

FW Tau was briefly observed with ALMA in Cycle 1 on 2013 December 2, using 27 
available 12 m antennas in an intermediate configuration (baseline lengths of 
17--460 m) and the Band 6 receivers under excellent conditions (0.6 mm of 
precipitable water vapor).  The correlator was configured to process four 
spectral windows in dual polarization, centered at 215, 217, 230.5, and 233 GHz 
(a mean frequency of 224 GHz, or $\lambda = 1.34$ mm), each with 128 coarse 
channels (15.625 MHz resolution) to maximize continuum sensitivity.  
Observations cycled between FW Tau and the nearby quasar J051002+180041 on 
$\sim$7 minute intervals, with additional visits to J0522-3627 and Ganymede for 
calibration purposes.  The total on-source time for the FW Tau field was 22 
minutes.

The raw ALMA visibilities were calibrated and imaged with the {\tt CASA} 
software package.  After phase correction using the water vapor radiometers, a 
system temperature correction, and initial flagging (which included the 
rejection of data from one antenna), the bandpass structure in each spectral 
window was corrected using observations of J0522-3627.  The absolute flux 
scaling was bootstrapped from observations of Ganymede.  Gain variations due to 
intrinsic changes in the array and atmosphere were determined from the 
monitoring of J051002+180041 and corrected.  The spectrally-averaged calibrated 
visibilities were Fourier inverted (assuming natural weighting), deconvolved 
with the {\tt CLEAN} algorithm, and restored with a $1\farcs45\times0\farcs75$ 
(P.A. = 131\degr) synthesized beam after a single iteration of phase-only 
self-calibration.  The RMS noise level in the resulting image is 28 $\mu$Jy 
beam$^{-1}$.

\section{Results}

\begin{figure}
\epsscale{1.2}
\plotone{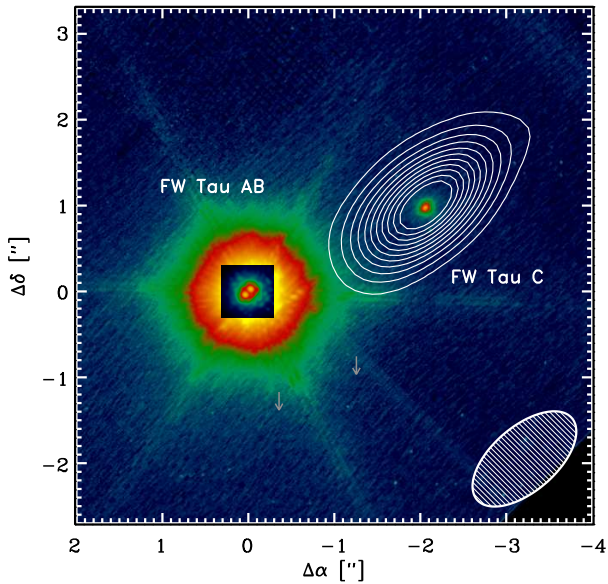}
\figcaption{\label{fig:image} ALMA 1.3\,mm continuum contours (shown at $5\sigma \approx 0.14$\,mJy beam$^{-1}$ intervals) are overlaid on a near-infrared ($K'$) image of the FW Tau system taken with the NIRC2 camera on the Keck-II telescope \citep{Kraus:2014lr}.  The mm-wavelength continuum emission is centered on the faint companion (FW Tau C or b, depending on its nature as a binary companion or planetary companion), and firmly detected at a peak $S/N$ of $\sim$60.  The ALMA synthesized beam dimensions are shown in the lower right corner.}
\end{figure}

Figure \ref{fig:image} shows contours of the ALMA 1.3 mm continuum emission overlaid on the Keck/NIRC2 $K^{\prime}$ image of the FW Tau components \citep[cf.,][]{Kraus:2014lr}.  Continuum emission at 1.3 mm is firmly detected (peak $S/N$ $\approx 60$) and centered on the faint companion, located 2\farcs28 $\pm$ 0\farcs05 or 330 $\pm$ 30 AU (P.A. = 296\degr $\pm$2\degr) from the FW Tau AB photocenter \citep[$\alpha$ = 04$^{\rm h}$29$^{\rm m}$29\fs71, $\delta$ = +26\degr16\arcmin52\farcs82, based on the ALLWISE astrometry and the proper motion estimated by][]{Kraus:2014lr}.  A Gaussian fit to the emission indicates that it is unresolved in the $\sim$200$\times$100\,AU beam and has an integrated $F_{\nu} = 1.78\pm0.03$ mJy (with an additional $\sim$10\%\ systematic uncertainty in the flux scale) as estimated from a Gaussian model of the visibilities.

This mm-wave continuum emission is faint enough to be entirely optically thin for any reasonable dust opacity \citep[e.g.,][]{Beckwith:1990uq}, so the flux measurement can be simply converted into a dust mass estimate.  Assuming a characteristic dust temperature of $\sim$10--20 K and a standard opacity of 2.3 cm$^{2}$ g$^{-1}$ at 1.3 mm, we infer $M_{\rm dust} \approx 1$--2 M$_{\oplus}$.  This measurement is entirely consistent with the unresolved, marginal detection of 850 $\mu$m emission ($F_{\nu} = 4.5\pm1.1$ mJy) from the FW Tau system by \citet[][]{Andrews:2005qf}; the corresponding spectrum scales like $F_{\nu} \propto \nu^{2.2\pm0.6}$, in excellent agreement with the mean value for disks in this wavelength range.  The Herschel fluxes from \citet[][]{Howard:2013qy} are also consistent with the far-IR SED that fits our ALMA flux, but do not provide a sufficient constraint on the SED shape to further constrain the dust temperature. The upper limit to the dust emission around the FW Tau AB binary, $F_{\nu} \le 90$ $\mu$Jy ($3\sigma$), is $\sim$30$\times$ lower than typical constraints, and corresponds to a dust mass limit of $\lesssim$0.1 $M_{\oplus}$.

The CO $J$=3$-$2 transition was not detected in the very wide channels (chosen to optimize continuum sensitivity).  The 3$\sigma$ upper limit on the peak flux is $\sim$3 mJy beam$^{-1}$ in a single 20 km s$^{-1}$-wide channel.

\section{Discussion}

\begin{figure}
\epsscale{1.2}
\plotone{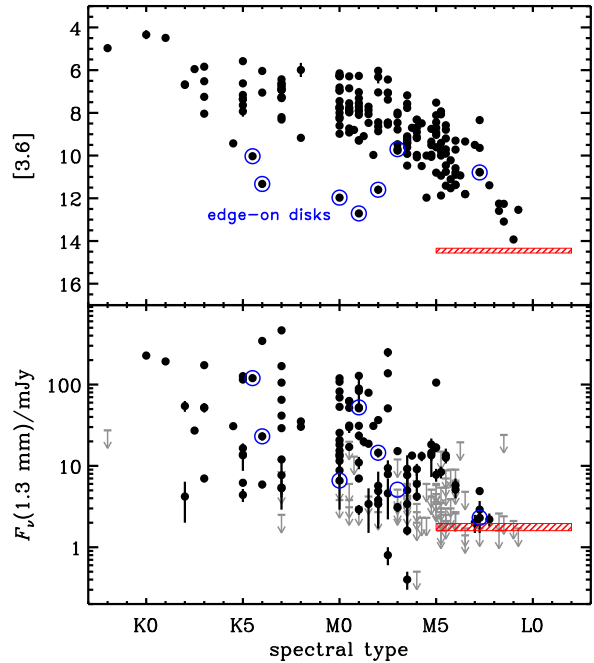}
\figcaption{\label{fig:comparison} A comparison of FW Tau C ({\it red, shaded bar} reflecting the uncertainties) and the population of disk-bearing stars and brown dwarfs in the $\sim$2\,Myr-old Taurus-Auriga star-forming region, using the empirical scaling of 3.6\,$\mu$m magnitudes and 1.3\,mm fluxes (3$\sigma$ upper limits are marked as gray arrows) with spectral type.  The Taurus-Auriga data were collected as described by \citet{Andrews:2013uq,Akeson:2014lr} and \citet{Luhman:2010cr}; the FW Tau C measurements were converted from the \citet{Kraus:2014lr} $L'$ photometry (using their conversion relation between the bands) and the ALMA data presented here. We use blue circles to denote the six stars in this sample that are either known or strongly suspected to have edge-on disk orientations, either from direct high-resolution imaging (HK Tau B, HV Tau C, 2MASS J04202144+2813491, and 2MASS J04381486+2611399; \citealt{Stapelfeldt:1998iv,Luhman:2007kl,Duchene:2010xt,Luhman:2010cr}) or a combination of an anomalously low $L_{*}$ (a $>$2$\sigma$ deviation from the mean luminosity at that spectral type), high $A_V$, or unusual optical/near-infrared SED morphology (IRAS 04260+2642, IRAS 04301+2608, and ITG 33A; \citealt{Andrews:2013uq}).}
\end{figure}

The ALMA data clearly show that all of the disk material in the system is associated solely with the faint companion, FW Tau C. Given the dust mass estimated from the observed 1.3\,mm flux ($\sim$1--3 $M_{\oplus}$) and its complementary signatures of disk accretion \citep{White:2001jf,Bowler:2014lr}, we have confirmed that the faint companion to FW Tau hosts one of the least massive primordial disks known to date. If that disk proceeds to form its own system of ``planetary" or ``satellite" companions, they will not achieve sufficient mass to become standard gas or ice giants. However, the mass is well matched to the total mass of compact systems of sub-Earth planets seen around field ultracool dwarfs (e.g., Kepler-42 bcd; \citealt{Muirhead:2012fe}. It is also interesting to note that the dust mass reservoir still exceeds the total sum of the Galilean satellites (0.066\,$M_{\oplus}$; \citealt{Showman:1999fj}) by just over an order of magnitude.

Expectations for the disk mass distribution for very low-mass primaries remain uncertain throughout the substellar and planetary mass regime, as there are only a handful of detections and many non-detections \citep[e.g.,][]{Scholz:2006lw,Schaefer:2009xi,Andrews:2013uq,Ricci:2014lr}. However, a comparison of the infrared and millimeter fluxes to a population of free-floating young counterparts could still provide context as to the nature of the disk host as either a brown dwarf tertiary companion (FW Tau C) or a PMC (FW Tau b). In Figure 2, we show how the 3.6\,$\mu$m magnitude \citep[converted from the $L'$ measurement reported by][]{Kraus:2014lr} and the new 1.3\,mm flux for the companion compare to the flux versus spectral type relations for disk-bearing stars and brown dwarfs in the Taurus-Auriga region \citep{Luhman:2010cr,Andrews:2013uq,Akeson:2014lr}. A shaded band is used for FW Tau C to denote the range of possible spectral types, limited at the upper end ($\sim$M5) by the presence of deep water absorption bands in the near-infrared spectrum \citep{Bowler:2014lr}.

At 3.6\,$\mu$m, FW Tau C is fainter than nearly all known substellar companions in the region, clearly in the regime of objects with spectral types later than M9 ($M < 10$\,$M_{Jup}$), though some edge-on disks also sit well below the median relation.  However, comparisons of its millimeter flux with the low-mass Taurus-Auriga population are more ambiguous. Most mm-wave surveys for brown dwarf disks were conducted with facilities that are substantially less sensitive than ALMA. The best available upper limits of those surveys approach the 1.3\,mm flux detected here for FW Tau C, so there are at least some brown dwarf disks with masses that could be quite a bit lower. The disk mass measured here is consistent with the range of possible values for isolated Taurus-Auriga brown dwarfs with spectral types later than $\sim$M5, assuming it has a similar age and formed in a similar process.

A more appropriate comparison would consider only companions, rather than the full population of isolated objects. Even a very low-mass companion might be able to retain a substantial disk via Bondi-Hoyle accretion from the disk or envelope of the primary, but free-floating objects do not have such a reservoir to build or maintain a comparable disk mass. Unfortunately, the number of substellar companions in nearby star-forming regions is quite low, and the number with component-resolved millimeter measurements is even lower, so such a comparison is not yet feasible. 

Trends observed for more massive stars might also apply, but it is unclear which processes are dominant in this specific case. A clear trend is seen in older regions for less massive free-floating objects to retain their disks longer \citep{Carpenter:2006hf}, perhaps even for several tens of Myr among brown dwarfs \citep{Riaz:2008uq,Reiners:2009aa}. Observations of (stellar) binaries in Taurus suggest that primaries usually dominate the disk mass budget \citep{Harris:2012oz}, though perhaps not out of proportion to the observed scaling of disk mass with primary mass \citep{Akeson:2014lr}, and the disks ultimately seem to have similar lifetimes \citep{Prato:1997qv,Daemgen:2012jl}. Furthermore, close binaries (such as the FW Tau AB pair, with $\rho \sim 15$ AU) appear to have sharply reduced disk frequencies even at early ages \citep{White:2001jf,Cieza:2009fr,Duchene:2010cj,Kraus:2012qe}, so it might not be surprising that the primary has cleared its disk quickly. The relative rate of disk dispersal is therefore an ambiguous feature in the classification of this system.

In the meantime, the most unambiguous way to determine the mass and geometry of FW Tau C will be to spectroscopically detect photospheric absorption features, either at higher spectral resolution or in the $\sim$1\,$\mu$m regime where veiling is minimized.  Disk mass determinations for more securely identified PMCs also could shed light on the distinction between companions formed via binary processes or planetary processes. In any case, the ALMA data presented here signals that we are entering a new regime of sensitivity that will ultimately enable the characterization of fundamental circum(sub-)stellar and circumplanetary disk properties.

\acknowledgements

The authors thank Gaspard Duch\^ene for an insightful discussion of the nature of the FW Tau companion, and the referee for a helpful critique of this paper. This paper makes use of the following ALMA data: ADS/JAO.ALMA\#2012.1.00989.S . ALMA is a partnership of ESO (representing its member states), NSF (USA) and NINS (Japan), together with NRC (Canada) and NSC and ASIAA (Taiwan), in cooperation with the Republic of Chile. The Joint ALMA Observatory is operated by ESO, AUI/NRAO and NAOJ. The NRAO is a facility of the NSF operated under cooperative agreement by Associated Universities, Inc.


\end{document}